# Point systems in Games for Health: A bibliometric scoping study

*Peter Kokol, University of Maribor; Faculty of Electrical Engineering and Computer Science*

*Koroška 46, 2000 Maribor, Slovenia*

*Peter.kokol@um.si*

**Aims:** Very few details about point systems used in games for health are reported in scientific literature.

**Methodology:** To shed some light on this topic a bibliometric study, analyzing the papers containing terms related to games for health and point systems was performed and a mini taxonomy was derived. The search string *game\* AND health AND (point\* OR score) AND system\** in a Scopus bibliographic database was used to produce the corpus. We limited the search to articles, reviews and conference papers written in English and to topics related to medical, health and social subjects. The corpus papers abstracts and titles were analysed by VOSviewer and a scientific landscape was generated. The landscape clusters were transformed into a mini – taxonomy.

**Results and conclusions:** The search resulted in a corpus consisting of 354 papers. The derived taxonomy contains three objects; video games, serious games and educational games. The biblimetric mapping and taxonomy revealed some interesting conclusions: (1) the video games have mostly negative effects on health, (2) the serious games might have both a direct positive health effects on users and also indirect effects by improved competencies of health professionals, and (3) the research is concerned not only to computer based educational games, but also to traditional table games and sporting games. Based on the derived taxonomy we can conclude that point systems should reward physical activity and healthy living style and punish sedentary activities.

**Category:** Other

# 1. INTRODUCTION

Various changes in health care systems like changing government's role in financing, service provision, changed regulation, new health doctrines, demographic changes (increased number of elderly people with chronic and other diseases and reduced functionality, reduced birth rate due to birth and infertility problems, increased number of divorces, chancing family structure, migration etc.), environmental changes, fast pace of life (which is closely linked with maximizing profits and focused on obtaining material goods), physical inactivity, substance abuse, unhealthy diets, etc they all have negative effect on lifestyles and health status of entire populations as well as individuals thus increasing the pressure on public health systems. This pressure is becoming worse with the emergence of complex and costly treatment possibilities. The information age and the fast development of information and communication technologies (ICT) can on one side contribute to the prevalence of sedentary and un-healthy live style, but can at the same time provide means for improving all areas of individual and public health. One of the many possibilities how to achieve the health style improvement are so called serious games or Games for Health (G4H).

Scoring systems are a key component of game mechanics, and provide a mechanism whereby players are rewarded with point value whenever they accomplish a task in the game. The growing complexity of scoring systems underlines the importance of determining the degree to which the design of a scoring system affects player satisfaction.

Scoring systems which rewarded players with point values are a key component of achieving the game goals and to motivate players. Lately, the scoring mechanism are becoming more and more complex thereafter its surprising that so few details of point system for games for health are reported. To shed some light on this topic we performed a bibliometric study, analysing the research literature production related to point systems in games for health.

# 2. METHODOLOGY

Bibliometrics analysis measures statistical properties of written materials like articles, conference papers and books [1]. One of the bibliometric approaches approaches is called bibliometric mapping [2]. It visualizes scientific literature production in form of maps (citation maps, term maps, collaboration networks, etc). In our study we used a popular bibliometric mapping software titled VOSviewer V1.6.6 (Leiden University, Netherlands) [3]. VOSviewer among other maps generates so called scientific landscapes, which express terms relatedness (proximity of terms), associations between terms (clusters) and importance of a term (term size). The VOSviewer mapping techniques are similar to those of multidimensional scaling [4]

## 2.1. Data set

Scopus (Elsevier, Netherlands) is the largest bibliographic database with rigorous indexing criteria and that was the main reason that we selected it and it alone for the publications harvesting. To find the most optimal search string we experimented with different keywords and various combinations among them; Finally we selected the following search string: *game\* AND health AND ((point\* OR score) AND system\*).* We limited the search to articles, reviews and conference papers written in English. The whole time period covered by Scopus was searched, however we limited the subjects to medicine, health and social topics. The search was performed on 14[th] of November 2017. The paper abstracts and titles were analysed by VOSviewer. A challenge in the generation of scientific landscapes is the selection of terms. Many stop word like the, is, are, and, of, between, etc.

are automatically removed by the VOSviewer text mining algorithm, terms like 'theory', 'approximation', 'dependence', 'study' 'correlation', 'baseline', 'period', 'possibility', 'calculation', 'significance', 'comparison', 'assumption', were removed manually. Additionally all terms with the occurrence less than 10 were ignored. To optimize the borders between clusters, the minimal cluster size was set to 8. All other VOS viewer parameters were left at default values.

## 3. RESULTS AND DISCUSSION

The search described above resulted in a corpus consisting of 354 publications. The VOSviewer analysis of the corpus resulted in the scientific landscape presented in Figure 1. It consists of three clusters shown in green, red and blue colour. Using the taxonomy development approach proposed by Nickerson, Varshney and Munterman [5], taking clusters as the units of interest resulted in the taxonomy shown in Table 1.

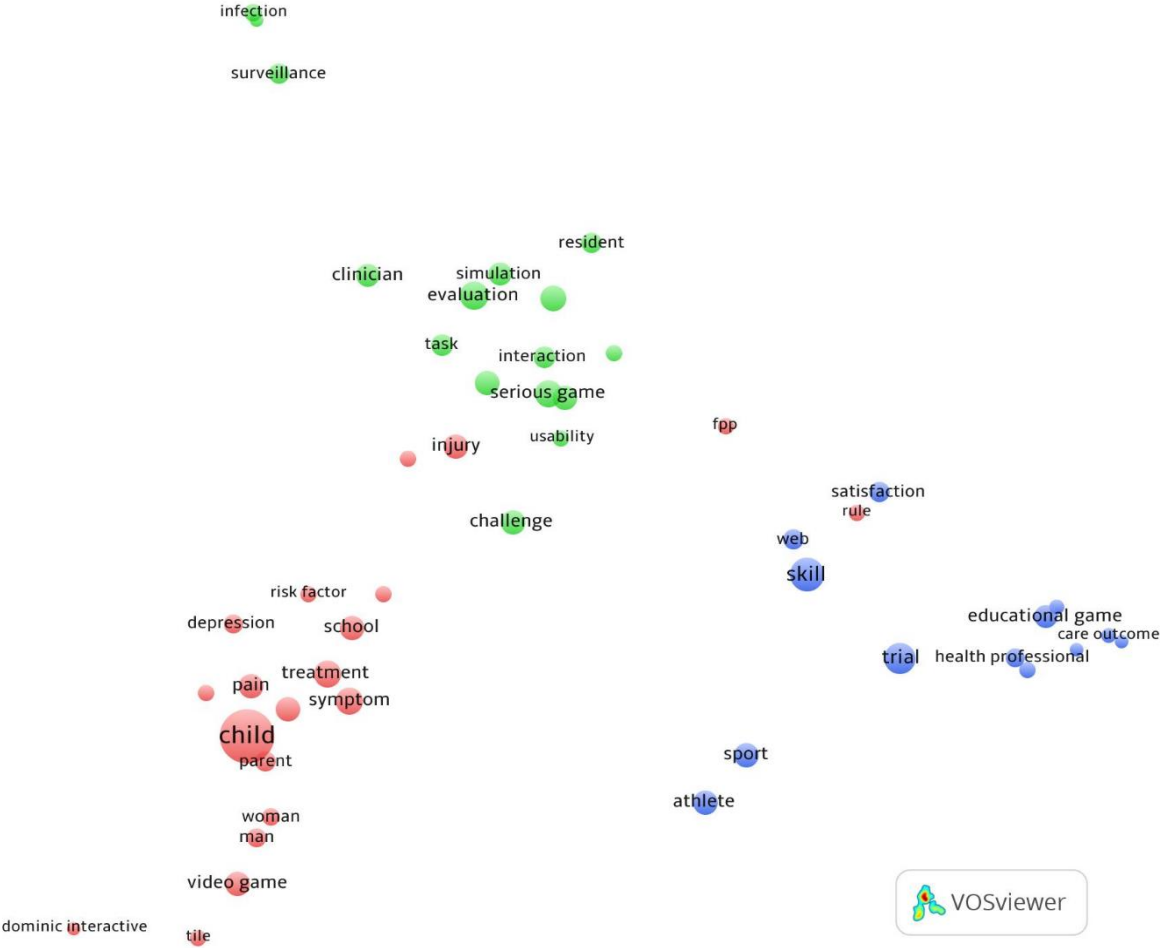

Figure 1. The scientific landscape for the point system in games for health

Table 1. The taxonomy of research literature related to games for health and point systems

| Topic | Representative terms | Effects from playing video games reported in most cited papers |
|---|---|---|
| Video games (red) | Video game, child, physical activity, dominic interactive, depression, risk factor, injury | Negative effects of video games on physical activity [6]. Internet gaming disorder [7]. |
| Serious games (green) | Serious game, simulation, challenge, task, interaction, usability, evaluation | Serious games use can improve health of patients [8,9] and serious games like simulation can improve the services of health professionals [10,11]. |
| Educational games (blue) | Educational game, health professional, sport, athlete | Use of ICT computer based educational games can improve learning of health professionals [12, 13] and students [14,15] |

Manual analysis of the most cited publications related to each topic of the taxonomy revealed some interesting observations about the research literature about point systems used in computer games for health:

- Nevertheless that we analysed publications related to point systems in computer games for health, there were very little details about point systems used.
- According to the point systems three different types of computer games emerged.
- It seems that the video games have mostly negative effects on health.
- The serious computer games might have both a direct positive health effects, however also indirect effects by improved competencies of health professionals acring for patients.
- The research in point systems is concerned not only to computer based games, but also to traditional table games and sporting games.

## 4. CONCLUSION

Based on the derived taxonomy and analysing the most cited papers we can recomend that point systems should reward physical activity and healthy living style and punish sedentary activities. In that manner games for health might be implemented as mobile applications which use the mobile phone accelerometer to track movement in various physical activities and award points based on intensity, distance or number of repetitions. Alternatively, the application may require from the user to reach predefined geographical points, make spictures on specific locations, read a "magic code" in various places, find a treasure, etc. Awarded point can then be used to advance to higher levels, "buy" equipment, acquire special powers and similar.

## 5. REFERENCES

1.Nicola De Bellis, *Bibliometrics and Citation Analysis*. Lanham, Maryland, Toronto, Plymouth, UK: The Scarecrow Press, Inc., 2009.

2.P M Alfonzo, T J Sakraida, and M Hastings - Tolmsa, "BIbliometrics: Visualizing the impact of nursing research," *Online Journal of Nursing Informatics*, vol. 18, no. 1, pp. 3 - 17, 2014.

3.F W Young and R M Hamer, *Multidimensional Scaling: History, Theory, and Applications*. Psychology Press: Florence, Kentucky, USA , 2013.

4.N J Van Eck and L Waltman, "Bibliometric mapping of the computational intelligence field," *International Journal of Uncertainty, Fuzziness and Knowledge-Based Systems*, vol. 15, no. 5, pp. 625-645, 2007.

5.R C Nickerson, U Varshney, and J Muntermann, "A Method for Taxonomy Development and its Application in Information Systems," *European Journal of Information Systems*, vol. 22, no. 3, pp. 336 - 353, 2013.

6.E B Kahn et al., "The effectiveness of interventions to increase physical activity: A systematic review," *American Journal of Preventive Medicine*, vol. 22, no. 4, pp. 73 - 107, 2002.

8.R O Guiterrez et al., "A telerehabilitation program by virtual reality-video games improves balance and postural control in multiple sclerosis patients," *NeuroRehabilitation*, vol. 33, no. 4, pp. 545 - 554, 2013.

10.R C Basole, D A Bodner, and W B Rouse, "Healthcare management through organizational simulation," *Decision Support Systems*, vol. 55, no. 2, pp. 552 - 583, 2013.

12.E A Aki et al., "Educational games for health professionals," *Cochrane Database of Systematic Reviews*, p. Article number CD006411, 2008.

14.S Rondon, F C Sassi, and C R Furquim De Andrade, "Computer game-based and traditional learning method: A comparison regarding students' knowledge retention," *BMC Medical Education*, vol. 13, no. 1, p. AN 30, 2013.

7.K A Faust and J J Prochaska, "Internet gaming disorder: A sign of the times, or time for our attention?," Addictive Behaviours, vol. 77, pp. 272-274, 2018.

9.J Jonsdottir et al., "Serious games for arm rehabilitation of persons with multiple sclerosis. A randomized controlled pilot study," Multiple Sclerosis and Related Disorders, vol. 19, pp. 25-29, 2018.

11.D Drummond et al., "Serious game versus online course for pretraining medical students before a simulation-based mastery learning course on cardiopulmonary resuscitation," European Journal of Anaesthesiology, vol. 34, no. 12, pp. 836-844, 2017.

13. L Sera and E Wheeler, "Game on: The gamification of the pharmacy classroom," Currents in Pharmacy Teaching and Learning, vol. 9, no. 1, pp. 155-159, 2017.

15. S Cardullo, L Gamberini, S Milan, and D Mapelli, "Rehabilitation Tool: A Pilot Study on A New Neuropsychological Interactive Training System," Studies in Health Technology and Informatics, vol. 2019, 2015.